# Finite image size effects on the characterization of magnetic domain patterns via magnetic force microscopy

Michael Vaka[1], Joey Ray[1], Misael Campos[2], Karine Chesnel[1,a]

[1]Department of Physics and Astronomy, BYU, Provo, UT, 84058, USA

[2]Department of Electrical Engineering, Princeton University, Princeton, NJ, 08544, USA

a)  Author to whom correspondence should be addressed: kchesnel@byu.edu

**Abstract**

Optimizing magnetic thin films for nanotechnologies often requires imaging nanoscale magnetic domain patterns via magnetic microscopy. The finite size of the image may however significantly affect the characterization of the observed magnetic states. We evaluated finite image size effects on the characterization of a variety of stripe and bubble domain patterns exhibited by ferromagnetic Co/Pt multilayers with perpendicular magnetic anisotropy, where the domain size (stripe width and bubble diameter) is around 100 nm. If the image size is too small, below ~ 5 µm, it may cause a significant underestimation of average domain size and overestimation of domain density by up to a factor 5 when reducing the image size from about 20 µm to about a 1 µm. Using a criterion based on how the excess density evolves with image size, we found that to obtain reliable statistical estimates of domain density and average domain size, the image needs to be large enough, and include at least about 100 stripes or about 2500 bubbles.

**Introduction**

Magnetic recording nanotechnologies heavily relies on the optimization of magnetic media. In the quest for ever-increasing storage capabilities, thin ferromagnetic films exhibiting perpendicular magnetic anisotropy (PMA) have been key materials as they allow achieving high density of nanoscale magnetic domains. [1-5] Thin Co/Pt and Co/Pd multilayers with PMA are such materials that exhibit high magnetic domain densities [6-8], and where the domain pattern can serve as a template for magnetic domain memory applications [9,10]. We found in previous studies on $[Co(x)/Pt(0.7nm)]_N$ that when the Co thickness is optimized to x ≈ 3 nm and the number of repeat optimized to N ≈ 20, the magnetic domain pattern present in the material at remanence can undergo a full transition from a stripe state to a lattice of bubble domains where the density of domain is drastically increased by a factor 10 or more [11-13]. To best characterize and control these morphological transitions, one needs to well visualize the magnetic domain patterns in the film at the nanoscale. [12,14] With spatial resolution down to ≈ 20-25 nm, magnetic force microscopy (MFM) is a common way to image nanoscale magnetic domain patterns in thin ferromagnetic films with PMA. However, the scanning nature of MFM often requires limiting the size of the image, typically in the range of 1 - 20 µm and finding a tradeoff between the scanning duration and spatial resolution. When quantifying physical features such as average domain size and domain density, finite size effects may significantly affect statistical estimates, especially when magnetic domains get massively cut at the edge of the image. Because the characterization of magnetic states and morphological magnetic transitions heavily rely on the accuracy of such estimates, it is crucial to correctly account for any existing finite image size effect.

**Methods**

To address the finite size effect question, we study here three MFM images with distinctive domain patterns: stripe, bubble and mixed patterns. These patterns were observed in [Co (3 nm) /Pt (0.7 nm)]$_N$ multilayers exhibiting PMA, where the Co thickness of 3 nm and the number of repeat *N* had been optimized to maximize domain densities, based on our previous studies [13,14]. The bubble and stripe patterns were achieved in the same material, for which *N* = 20, by applying an out-of-plane magnetic field of different magnitudes: 4 kOe for the stripe pattern and 6 kOe for the bubble pattern. The mixed state was obtained on a sample with *N* = 18 by bringing the material to remanence after applying a field of 7 kOe. The MFM images were collected with an in-situ field on a Nanoscope V Dimension 3100 instrument. To obtain reliable statistical results, large images up to 40 x 40 µm$^2$ for the stripe pattern and 20 x 20 µm$^2$ for the bubble and mixed patterns were collected. For the stripe, bubble, and mixed patterns, the scan rates were 0.254 Hz, 0.509 Hz, and 0.500 Hz per line, respectively. The spatial resolution, defined here by the pixel size was 9.8 nm (40 µm / 4096 pixels), 19.5 nm (20 µm / 1024 pixels), and 39 nm (20 µm / 512 pixels), respectively. These large images were binarized, partitioned into square blocks, and statistics were computed over all the blocks.

To estimate domain densities and domain sizes, the MFM images were binarized using MATLAB. We designed an algorithm that uses a *binarize* function [15] with adaptive thresholding [16] based on Gaussian statistics, followed by morphological line erosion and disk opening. The binarization was optimized using the sensitivity, neighborhood size, morphological line length and line angle, and morphological disk radius parameters. The 2D correlation [17] was maximized using the *Pattern Search* optimizer [18]. The correlation values achieved in the stripe, bubble, and mixed pattern were 0.906, 0.855, and 0.799 respectively. This demonstrates sufficiently good binarization for the given resolutions. The resulting binarized images represent local magnetization where white and black respectively correspond to aligned and reversed magnetization direction with respect to the applied out-of-plane magnetic field. The binarized images were then partitioned into square blocks of length *L*, with *L* successively taking the values of 20, 10, 5, 2.5 and 1.25 µm.

Once the binarization was completed, we extracted domain properties by using the MATLAB's *Region Property* function [19] with the number of connected regions and area of these regions being the output properties for white and black domains, respectively. The domain density of a given color was estimated by dividing the total number of domains of that color by the block area and rescaling the result per 100 $\mu m^2$. This operation was carried on each individual block of size *L*. The individual densities were averaged over all the blocks, for a given *L*. The density plots show these average domain densities along with their standard deviation, as a function of *L*. Domain area distributions (averaged over all the blocks) are also shown for each given *L*. In the presented histograms, the bin size is 0.01 $\mu m^2$ for the stripe pattern and 0.001 $\mu m^2$ for the bubble and mixed patterns. The average area plots show the average domain area, along with its standard deviation, for each given $L$. These averages excluded outliers. The interpolation for the comparative density plots was carried out using a pchip algorithm.

**Results**

Results for a predominant stripe pattern are displayed in Fig.1. The close-up view in Fig.1a taken from a 40 x 40 µm$^2$ MFM image shows a domain pattern mostly made of aligned stripes with sparse forks and a few trapped bubbles. The extracted densities, plotted in Fig.1b, show the same trend for the reverse (white) and aligned (black) domains. Starting from the smallest partition size L, the domain density $\rho$ rapidly drops to eventually plateau when L increases. For the reversed (white) domains, the average density $\rho_w$ decreases from ~ 760 domains/ 100 µm² at L = 1.25 µm down to ~ 148 domains/ 100 µm² at L = 40 µm. The plateauing value is interpreted as the true density value $\rho_w$ one would measure in the absence of finite scan size effects (infinite image). Likewise, the average density $\rho_b$ for the aligned (black) domains decreases from 525 domains/ 100 µm² at L = 1.25 µm down to 3 domains/ 100 µm² at L = 40 µm. In this pattern, the presence of forks and bubbles causes $\rho_w > \rho_b$ with the difference $\rho_w - \rho_b$ eventually reaching 145 domains/ 100 µm².

The distribution of individual domain areas in Fig.1c, shows a wide spread of areas from 0 to 1 µm². Depending on the partition size L, additional peaks appear at random locations on top of the uniform base distribution. From these area distributions, an averaged domain area was extracted in Fig 1d. The average domain area gradually increases when L increases, growing from 0.06 µm² at L = 1.25 µm to 0.32 µm² at L = 40 µm, a drastic increase by a factor 5.3. This trend is consistent with the density trend, illustrating the effect of the finite scan size, resulting in stripes being artificially ended at the edges of the image.

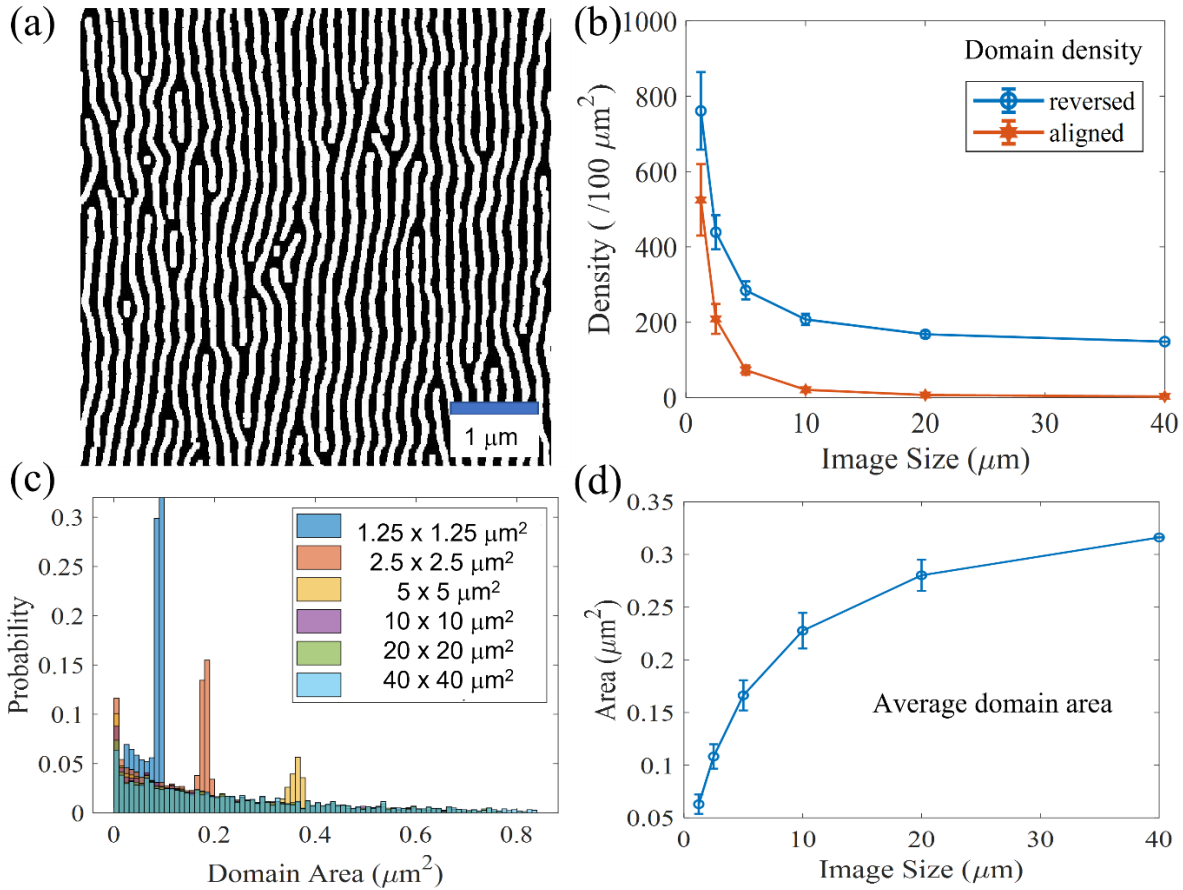

FIG. 1. Results for a predominantly **stripe pattern**. (a) Close-up view from a 40x40 µm² MFM image (binarized); (b) Density of white domains $\rho_w$ and black domains $\rho_b$, as function of the image size L it is estimated on; (c) Distribution of domain areas for the reversed (white) domains for various partition sizes L. The bin size is 0.01 µm²; (d) Averaged domain area as function of the image size L.

Results for a mixed stripe-bubble pattern are displayed in Fig.2. The close-up view in Fig.2a taken from a 20 x 20 µm² MFM image shows a mix of distinct aligned stripes and chains of bubbles. The extracted black and white domain densities, $\rho_b$ and $\rho_w$, plotted in Fig.2b, show the same trend with *L*. This trend mimics the trend previously observed in the case of stripe pattern, except, in the mixed state, the decrease is not as steep and the difference $\rho_w - \rho_b$ is more pronounced, now reaching 716 domains/ 100 µm² at *L* = 20 µm. Also, the plateauing value for $\rho_w$ is much higher, now around 760 domains/ 100 µm², while the plateauing value for $\rho_b$ remains relatively low at 45 domains/ 100 µm².

The distributions of white domain areas, in Fig.2c, shows a nearly Gaussian distribution that peaks at 0.0065 µm² for all the partition sizes *L*. We attribute this peak to the large number of bubble domains. The peak area of 0.0065 µm² corresponds to an average bubble radius of 45 nm (bubble diameter 90 nm), consistent with earlier findings for N=18 [13]. From these area distributions, an average domain area, averaged over all (bubble and stripe) domains, was extracted and plotted in Fig 2d. The average domain area gradually increases when *L* increases, growing from 0.031 µm² at *L* = 1.25 µm to 0.047 µm² at *L* = 40 µm. This 52% increase is relatively moderate compared to the extent of the distributions, with standard deviations as large as $\pm 0.013 \ \mu m^2$. Consistent with the density trend, this increase in average domain area is essentially due to the artificial shortening of the stripes being cut at the edge of the image. However, the consistent peak position in the area distribution indicates that the average bubble size is not affected by the image size *L* (as long as *L* is larger than the bubble size, a condition that is largely satisfied here).

Results for a predominant bubble pattern are displayed in Fig.3. The close-up view in Fig.3a taken from a 20 x 20 µm² MFM image shows a lattice of bubbles. The extracted densities, $\rho_b$ and $\rho_w$, plotted in Fig.3b, show the same trend, where the domain density $\rho$ decreases and eventually plateau when *L* increases. This trend mimics the trend previously observed in the case of stripe and mixed patterns, except, in the bubble state, the difference $\rho_w - \rho_b$ is even more pronounced, reaching up to 1898 domains/ 100 µm² at *L* = 20 µm, with plateauing values of 1907 domains/ 100 µm² for $\rho_w$ and only 9 domains/ 100 µm² for $\rho_b$. This large discrepancy between white and black densities reflects the predominance of white bubble over a black background.

The distribution of individual domain areas, in Fig.3c, peaks at ~ 0.012 µm² for all the partition sizes *L*. The peak area of 0.012 µm² corresponds to an average bubble radius of 62 nm (bubble diameter 124 nm), consistent with earlier findings for *N* = 20 [13]. The area distribution is narrower compared to the stripe and mixed patterns, with a standard deviation smaller than $\pm 0.003 \ \mu m^2$ in the case of the bubble pattern. Also, the average value is much closer to the peak value, indicating that most domains are bubbles of similar size. The extracted average domain area, averaged over all domains, is plotted in Fig 3d. The average domain area slightly increases when *L* increases, growing from 0.017 µm² at *L* = 1.25 µm to 0.022 µm² at *L* = 20 µm. This 28 % increase is much smaller compared to the case of stripe patterns, due to a smaller proportion of long stripes in the bubble pattern, resulting in limited cutting effects at the edge of the image.

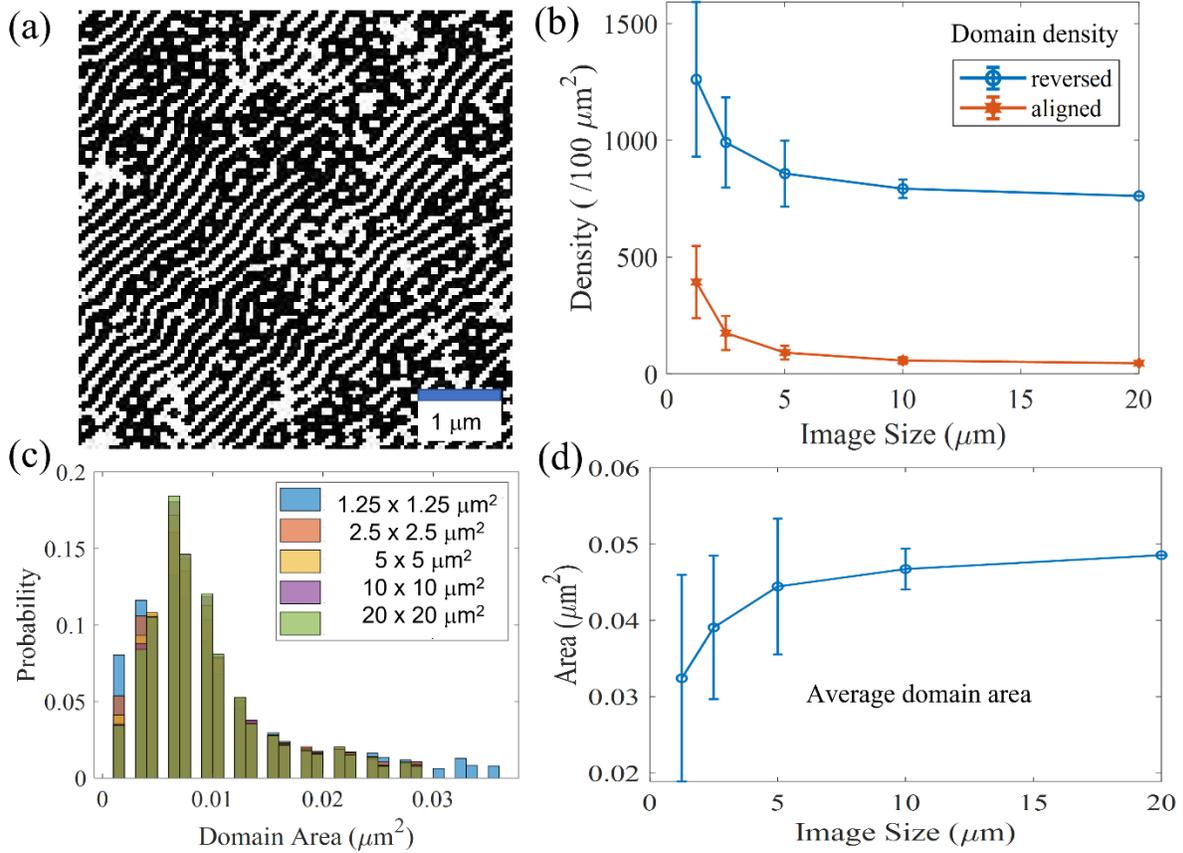

FIG. 2. Results for a **mixed stripe-bubble pattern**. (a) Close-up view from a 20 x 20 µm² MFM image (binarized); (b) Density of white domains $\rho_w$ and black domains $\rho_b$, as function of the image size $L$ it is estimated on; (c) Distribution of domain areas for the reversed (white) domains for various partition sizes $L$. The bin size is 0.001 µm²; (d) Averaged domain area as function of the image size $L$.

**Discussion**

Our data shows consistent finite scan size effects and suggests that the extent of these effects may depend on the morphology of the domain patterns. To compare the extent of the finite scan size effects in the various morphologies, we plotted in Fig.4 the density $\rho_w$ for the stripe, mixed and bubble patterns against each other. The absolute density plot in Fig.4a illustrates the gradual increase in density when evolving from stripes to bubbles, with the asymptotic density $\rho_w$ increasing from 148 domains/ 100 µm² in the stripe case, to 761 domains/ 100 µm² in the mixed case, and to 1898 domains/ 100 µm² in the bubble case. Comparing these densities once normalized to their asymptotic value, in Fig.4b, shows that the finite scan size effect is however much weaker for the bubble pattern compared to the stripe pattern. As summarized in Table 1, the density overestimation (artificial excess) $\Delta\rho$ caused by finite scan size only reaches 28% for the bubble pattern compared to 412% for the stripe pattern when $L$ = 1.25 µm.

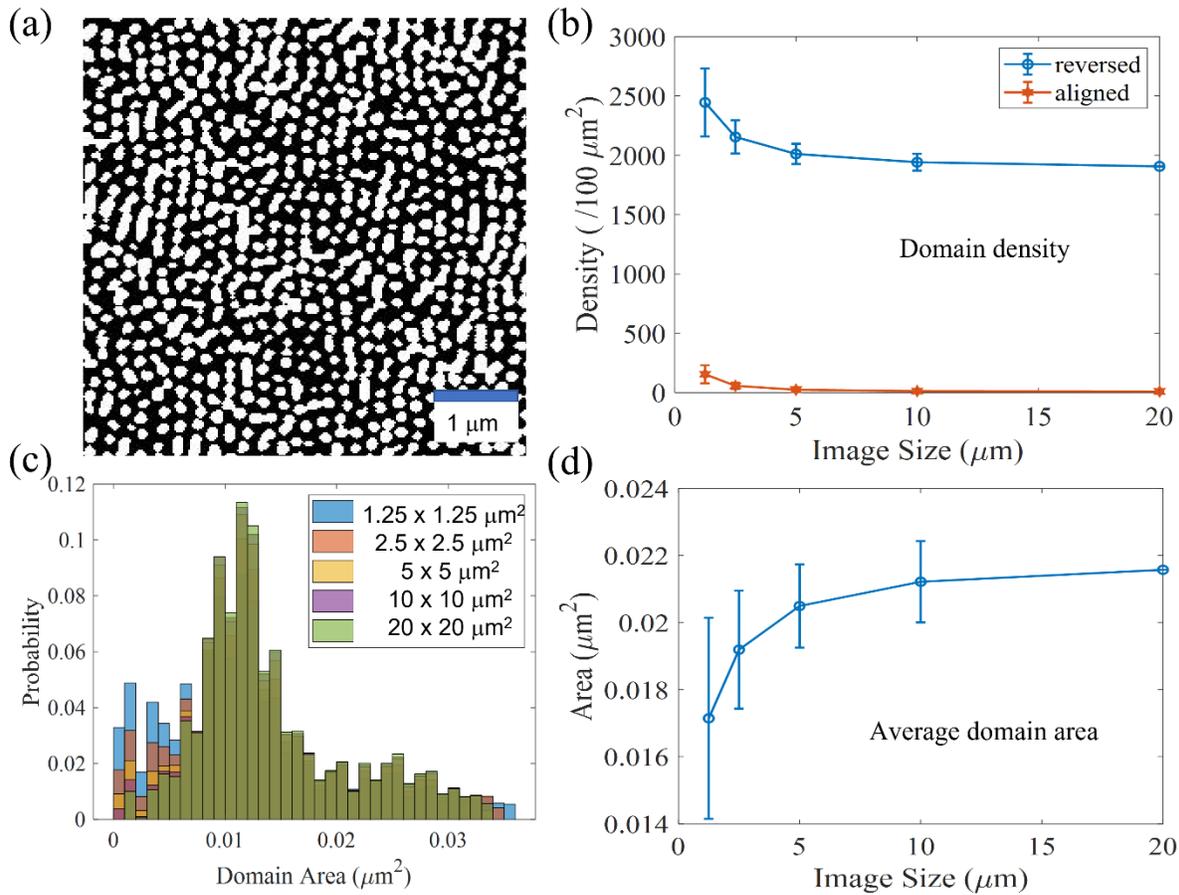

FIG. 3. Results for a predominantly **bubble pattern**. (a) Close-up view from a 20 x 20 µm² MFM image (binarized); (b) Density of white domains $\rho_w$ and black domains $\rho_b$, as function of the image size $L$ it is estimated on; (c) Distribution of domain areas for the reversed (white) domains for various partition sizes $L$. The bin size is 0.001 µm²; (d) Averaged domain area as function of the image size $L$.

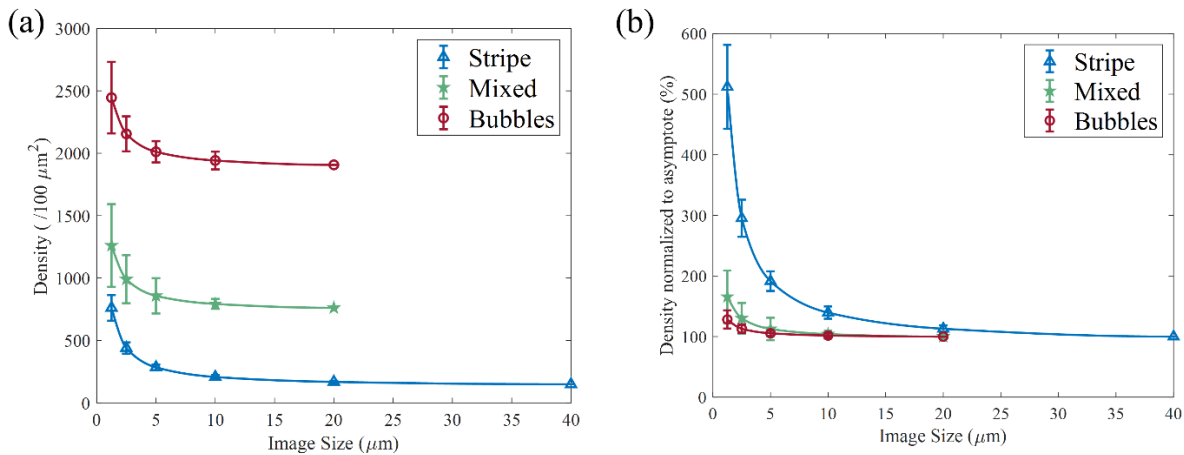

FIG. 4. Comparative plots for the stripe, mixed and bubble states. (a) Density $\rho_w$ of reversed domains as function of the image size $L$ it is estimated on, with interpolated curves; (b) Same density curves, normalized to their asymptotic value for comparison purposes.

From this data, we extracted, in Table 1, cutoff minimum image sizes using various criteria, based on limiting the excess density $\Delta\rho$ to 10% or to 1 %. Using the 1% criterion, the data suggests that the minimum image size to obtain reliable accurate statistics for our bubble patterns is around 13 x 13 $\mu m^2$. With bubble sizes in the range of 100-120 nm, and distance between bubbles around 250 nm, this corresponds to about 2500 bubbles per image or ~ 50 bubbles in any direction if assuming a regular hexagonal lattice of bubbles. For the stripe pattern, the 1% criterion leads to a minimum image size of around 35 x 35 $\mu m^2$. With the stripe width matching the bubble diameter of ~100-125 nm and a black-white stripe period in the range of 200-250 nm, it corresponds to about 140 aligned stripes (of same color) per image.

|  | *Excess density $\Delta\rho$ (%) at various L* | | | | *Cutoff (minimum) image size ($\mu m$)* | |
|---|---|---|---|---|---|---|
| *Morphology* | 10 $\mu m$ | 5 $\mu m$ | 2.5 $\mu m$ | 1.25 $\mu m$ | $\Delta\rho < 10$ % | $\Delta\rho < 1$ % |
| Bubbles | 1.8% | 5.5% | 13% | 28% | 3.2 | 12.9 |
| Mixed | 4.1% | 13% | 30% | 66% | 6.2 | 15.6 |
| Stripes | 39% | 91% | 195% | 412% | 23 | 35 |

Table 1. Selected values for the excess density and cutoff image sizes for the various morphologies: bubble, mixed and stripe states, where the bubble diameter and stripe width are both around 100 nm. The excess density $\Delta\rho$, here listed for various image sizes $L$, is normalized to the asymptotic density value and provided in %, (also see Fig. 4b). The cutoff image size is estimated using various criteria: the normalized excess density $\Delta\rho < 10$ % and 1 %.

Additionally, the data suggests that neither the scanning rate nor the pixel size affect the domain density estimate, as long as the Nyquist sampling criterion is satisfied, i.e., there are enough pixels per domain. If the pixel size is too big with respect to domain size, the Nyquist criterion may not be satisfied for all domains, producing poor quality image and affecting the accuracy of the binarization process, and affecting the estimation of domain densities and domain average size. Here, the Nyquist criterion is largely satisfied for all the presented images, with pixel sizes smaller than 40 nm and down to 10 nm, and domain sizes larger than 100 nm, giving a minimum of 5 pixels and up 100 pixels to define each individual bubble domain. With the Nyquist criterion being satisfied, the data suggests that the domain density and average domain size are not affected by the pixel size, but mainly by the total image size.

**Conclusion**

Our study shows that when imaging magnetic domain patterns, the finite size of the image can significantly affect the quantitative characterization of magnetic states and associated magnetic transitions. We conducted our study on thin Co/Pt multilayers with perpendicular magnetic anisotropy, exhibiting magnetic domain patterns of various shapes, from stripes to bubbles, depending on the applied magnetic field history. We found that the scan rate and pixel size did not affect our statistics, as long as the pixel size remains small enough to satisfy the Nyquist sampling criterion. However, we found that to obtain reliable statistical estimates of quantities such as domain densities and average domain sizes, the image size needs to be large enough, and even larger for stripes patterns compared to bubble patterns. We estimated cutoff minimum sizes based on the density excess relative to asymptotic values. If limiting the excess density to 1%, we found that in the case of a bubble pattern, the image needs to include at least about 2500 bubbles, or about 50 bubbles in any direction when in a close packed lattice of bubbles. For 100 nm bubbles, this would translate into minimum image size of about 10 x 10 $\mu m^2$. In the case of a stripe pattern, the image needs to include at least about 100 aligned stripes of same color, using that same criterion. For 100 nm wide stripes (200 nm period), this translates into 20 x 20 $\mu m^2$ minimum image size. If the image size is smaller than these suggested cutoff sizes, one should account for finite size effects, resulting in a significant underestimation of the domain size and overestimation of the domain density which may end up being several 100 % off.

**Acknowledgment**

We thank the group of Prof. Olav Hellwig from TU Chemnitz and HZDR in Germany, for sample supply. This research was supported by the NSF REU award #2051129 and the College of Mathematical and Physical Sciences (CPMS) at Brigham Young University (BYU).

**Data availability**

The data that support the findings of this study are available from the corresponding author upon reasonable request.